\documentclass[preprint,aps,prd,amsmath,amssymb]{revtex4}
\usepackage{graphicx}
\usepackage{color}
\newcommand{\beq}{\begin{eqnarray}}
\newcommand{\eeq}{\end{eqnarray}}

\newcommand{\bmp}{\noindent\begin{minipage}{16cm}}
\newcommand{\emp}{\end{minipage}\vskip 7mm} 

\usepackage{graphicx}
\usepackage{dcolumn}
\usepackage{bm}
\usepackage{amsmath}
\usepackage{amsfonts}
\usepackage{bbm}
\usepackage{subfigure}

\def\drawbox#1#2{\hrule height#2pt
        \hbox{\vrule width#2pt height#1pt \kern#1pt
              \vrule width#2pt}
              \hrule height#2pt}

\def\Asym#1#2{\vcenter{\vbox{\drawbox{#1}{#2}
              \kern-#2pt 
              \drawbox{#1}{#2}}}}

\begin{document}
\title{{\large Towards Working Technicolor}\\{\small  Effective Theories and
    Dark Matter}}
\author{Sven Bjarke {\sc Gudnason}}
\email{gudnason@nbi.dk}
\author{Chris {\sc Kouvaris}}
\email{kouvaris@nbi.dk}
\author{Francesco {\sc Sannino}}
\email{sannino@nbi.dk}
\affiliation{The Niels Bohr Institute, Blegdamsvej 17, DK-2100 Copenhagen \O, Denmark }

\begin{abstract}
A  fifth force, of technicolor type, responsible for breaking the
electroweak theory is an intriguing extension of the Standard
Model. Recently new theories have
 been shown to feature walking dynamics for
a very low number of techniflavors and are not ruled out by
electroweak precision measurements.
We identify the light degrees of freedom and construct the associated low energy effective theories. These can be
used to study signatures and relevant processes in current and future
experiments. In our theory the technibaryons are pseudo Goldstone bosons and their masses arise
via extended technicolor interactions. There are hypercharge assignments for the techniquarks which renders one of the
technibaryons electrically neutral. We
investigate the cosmological implications of this scenario and
 provide a component of dark matter.
\end{abstract}


\maketitle

\section{Introduction}
A dynamical mechanism behind the breaking of the electroweak
theory is a very intriguing possibility. A new strong
force is postulated to drive such a mechanism. The Large
Hadron Collider (LHC) experiment at CERN is going to probe
directly the sector associated to the breaking of the electroweak
theory and hence will be able to shed light on this force. Nature
has already shown to privilege such a mechanism which takes place
in ordinary superconductivity as well as in the spontaneous
breaking of chiral symmetry in quantum chromodynamics (QCD).

Earlier attempts using QCD-like technicolor \cite{TC}
have been ruled out by precision measurements
\cite{Peskin:1990zt}. Besides, one has also to face the problem of
mass generation which typically is provided by extended technicolor (ETC)
interactions and thus leads to large flavor changing neutral currents.
Recently it has been shown that one can construct viable theories
explaining the breaking of the electroweak theory dynamically
\cite{Sannino:2004qp,Dietrich:2005wk,Dietrich:2005jn,Sannino:2005dy,Evans:2005pu}
while not being at odds with electroweak precision measurements.
In the recently proposed theories technimatter transforms according to
a higher dimensional representation of the new gauge group.
By direct comparison with data it turns out that the preferred representation is the
two index symmetric \cite{Sannino:2004qp}. The simplest theory of
this kind is a two technicolor theory. In this case the two index
symmetric representation corresponds to the adjoint. Remarkably these
theories are already near conformal for a very small number of
techniflavors. Further properties of higher dimensional representations have been also explored in \cite{Christensen:2005cb}.  In \cite{Dietrich:2005jn,Evans:2005pu,Dietrich:2005wk}
the reader can
find a summary of a number of
salient properties of the new technicolor theories as well as a comprehensive review of the walking properties with references to the literature. We also note that near the conformal window \cite{Sundrum:1991rf,
{Appelquist:1998xf}} one of the relevant electroweak parameters ($S$) is smaller than expected in perturbation theory. This observation is further supported by very recent analysis \cite{Hong:2006si,{Harada:2005ru}}.

In this paper we examine the phenomenological implications of the
technicolor theory
with two techniquarks transforming according to the adjoint
representation of $SU(2)$. This theory has a $SU(4)$ quantum global
symmetry
which breaks spontaneously to $SO(4)$. Of the nine Goldstone bosons,
three are eaten by the electroweak gauge bosons while the remaining
ones carry
nonzero technibaryon number, which is associated with one of the
diagonal generators of $SU(4)$. These technibaryons must acquire a
mass from some, yet unspecified, theory at a higher scale. Since we
assume a bottom up approach we postpone the problem of producing the
underlying theory providing these masses, but we expect it to be
similar to the ETC type theory proposed in \cite{Evans:2005pu}. If the
technibaryon number is left intact by the ETC interactions
the lightest technibaryon (LTB) is stable and the
hypercharge assignment can be chosen in a way that the LTB is also
electrically neutral.

In the first part of the paper, we provide the associated linear and
non-linear effective theories. The latter can be used to make specific
quantitative predictions for the Large Hadron Collider
 as well as Linear Collider (LC).

If the technibaryon number is conserved the LTB is stable and it can
be made electrically neutral while its mass is expected
to be of the order of the electroweak scale. It has, hence, many
features required of a dark matter component. Following references \cite{Nussinov:1985xr,Barr:1990ca} we
have calculated
the contribution of this particle to dark matter and we found that it
can account for the whole dark
matter density. We should emphasize that in our calculation we
took under consideration the overall electric neutrality of the
matter in the universe as well as the thermal equilibrium conditions
and the sphaleron processes. There are no parameters to tune in
order to get the right technibaryon density other than the mass of
the neutral particle. Note that if dark matter is homogenously distributed in our galaxy our component cannot be
the dominant contribution to dark matter \cite{Akerib:2005kh} but can constitute part of it. However, it seems that the dark matter distribution in the Universe is not yet exactly determined \cite{Ma:2003cq} and depends crucially on the type and number of components.

We have made a preliminary investigation relative to the single step unification problem for the SM couplings and our technicolor coupling and found that they do not unify. This result is perhaps not too surprising since one expects new gauge theories to emerge before the typical unification scale and besides there could be multiple step unifications.

To date it is unclear if supersymmetry will ever play
a role in Nature. If supersymmetry is not discovered at the electroweak scale it can still emerge at much higher energies.
The Standard Model (SM) and the new strong
force will then become supersymmetric at this new higher scale. A
feature of this scenario, proposed long
ago by Dine, Fischler and Srednicki \cite{Dine:1981za}, is the
fact that these extensions of the Standard Model lend a natural solution to the so called $\mu$ problem of the minimal supersymmetric standard model (MSSM). If we adopt this idea we find that the our new strong force can be
extended to ${\cal N}=4$ super Yang-Mills by adding the missing scalars and suitably adjusting all of the interactions among the matter fields. We define $m$ as the
mass-scale above which the Higgs sector (i.e. now the technicolor
sector) of the theory becomes ${\cal N}=4$. The
corresponding coupling constant freezes above $m$ since this
sector of the theory is conformal. This would make our theory an
even better candidate for walking technicolor.
\section{The Model}
The new dynamical sector underlying the Higgs mechanism we
consider is an $SU(2)$ technicolor gauge group with two adjoint
technifermions. The theory is asymptotically free if the number of
flavors $N_f < 2.75$.

To estimate the critical coupling for chiral symmetry breaking we
required that the anomalous dimension of the quark mass operator
must satisfy the relation $\gamma(2-\gamma)=1$ \cite{Cohen:1988sq}. This yields
$\alpha_c  \simeq \frac{\pi}{3N}$.
The critical value of the number of flavors which gives this fixed
point value is $
 N_f^c \simeq 2.075$ \cite{Sannino:2004qp,Evans:2005pu}.

Since we consider adjoint Dirac fermions, the critical
number of flavors
is independent of the number of colors \cite{Evans:2005pu}. We expect that the
theory will enter a conformal regime unless the coupling rises
above the critical value triggering the formation of a fermion
condensate. Hence a $N_f=2$ theory is sufficiently close to the
critical number of flavors $N_f^c$. This makes it
a perfect candidate for a walking technicolor theory.

Although the critical number of flavors is independent of the number
of colors the electroweak precision measurements do depend on
it. Since the lowest number of colors is privileged by data
\cite{Dietrich:2005wk,{Dietrich:2005jn}} we choose the
two-technicolor theory.

Then the two adjoint fermions may be written as
\beq T_L^a=\left(  \begin{array}{c} U^{a} \\D^{a} \end{array}\right)_L ,
\qquad U_R^a \ , \quad D_R^a \ ,  \qquad a=1,2,3 \ ,\eeq
with $a$ the adjoint color index of $SU(2)$. The left fields are arranged in
three doublets of the $SU(2)_L$ weak interactions in the standard
fashion. The condensate is $\langle \bar{U}U + \bar{D}D \rangle$ which
breaks correctly the electroweak symmetry.

This model as described so far suffers from the Witten topological
anomaly \cite{Witten:fp}. An $SU(2)$ gauge theory must have an
even number of fermion doublets to avoid this anomaly. Here there
are three extra electroweak doublets added to the Standard Model
and we need to add one more doublet. Since we do not wish to
disturb the walking nature of the technicolor dynamics, the
doublet must be a technicolor singlet \cite{Dietrich:2005jn}. Our
additional matter content is essentially a copy of a standard
model fermion family with quarks (here transforming in the adjoint
of $SU(2)$) and the following lepton doublet \beq {\cal L}_L =
\left( \begin{array}{c} N \\ E \end{array} \right)_L , \qquad
N_R \ ,~E_R \ . \eeq
In general, the gauge
anomalies cancel using the following generic hypercharge
assignment
\begin{align}
Y(T_L)=&\frac{y}{2} \ ,&\qquad Y(U_R,D_R)&=\left(\frac{y+1}{2},\frac{y-1}{2}\right) \ , \\
Y({\cal L}_L)=& -3\frac{y}{2} \ ,&\qquad
Y(N_R,E_R)&=\left(\frac{-3y+1}{2},\frac{-3y-1}{2}\right) \ ,
\end{align}
where the parameter $y$ can take any real value. In our notation
the electric charge is $Q=T_3 + Y$, where $T_3$ is the weak
isospin generator. One recovers the SM hypercharge
assignment for $y=1/3$. In \cite{Evans:2005pu}, the SM
 hypercharge has been investigated in the context of an extended
technicolor theory. Another interesting choice of the hypercharge
is $y=1$, which has been investigated, from the point of view of
the electroweak precision measurements, in
\cite{Dietrich:2005jn,Dietrich:2005wk}. In this case
\begin{eqnarray}
Q(U)=1 \ , \quad Q(D)=0 \ , \quad Q(N)=-1 \ , \quad {\rm and }\quad Q(E)=-2 \ , \quad {\rm with} \quad y=1 \ .
\end{eqnarray}
Notice that in this particular hypercharge assignment, the $D$
technidown is electrically neutral. Since we have two Dirac fermions in the adjoint
representation of the gauge group the global symmetry is $SU(4)$.
In practice our technicolor sector has the same fermionic matter content as
that of ${\cal
  N}=4$ super Yang-Mills. To discuss the symmetry properties of the
theory it is
convenient to use the Weyl base for the fermions and arrange them in the following vector transforming according to the
fundamental representation of $SU(4)$
\beq Q= \begin{pmatrix}
U_L \\
D_L \\
-i\sigma^2 U_R^* \\
-i\sigma^2 D_R^*
\end{pmatrix},\eeq
where $U_L$ and $D_L$ are the left handed techniup and
technidown respectively and $U_R$ and $D_R$ are the corresponding
right handed particles. Assuming the standard breaking to the maximal diagonal subgroup, the $SU(4)$ symmetry breaks spontaneously
down to $SO(4)$. Such a breaking is driven by the following condensate
\beq \langle Q_i^\alpha Q_j^\beta \epsilon_{\alpha \beta} E^{ij} \rangle =-2\langle \overline{U}_R U_L + \overline{D}_R D_L\rangle \ ,
\label{conde}
 \eeq
where the indices $i,j=1,\ldots,4$ denote the components
of the tetraplet of $Q$, and the Greek indices indicate the ordinary
spin. The matrix $E$ is a $4\times 4$ matrix defined in terms
of the 2-dimensional unit matrix as
 \beq E=\left(
\begin{array}{cc}
0 & \mathbbm{1} \\
\mathbbm{1} & 0
\end{array}
\right) \ .
\eeq
Following the notation of Wess and Bagger \cite{WessBagger} $\epsilon_{\alpha \beta}=-i\sigma_{\alpha\beta}^2$ and
$\langle
 U_L^{\alpha} {{U_R}^{\ast}}^{\beta} \epsilon_{\alpha\beta} \rangle=
 -\langle  \overline{U}_R U_L
 \rangle$. A similar expression holds for the $D$ techniquark.
The above condensate is invariant under an $SO(4)$ symmetry. The
easiest way to check that an $SO(4)$ symmetry remains intact is by
going to the following base
\begin{equation}
U_L = \frac{\lambda_1+ i\lambda_2}{\sqrt{2}} \ ,\quad  \epsilon U_R^* = \frac{\lambda_1 - i\lambda_2}{\sqrt{2}} \ ,\quad D_L = \frac{\lambda_3+ i\lambda_4}{\sqrt{2}} \ ,\quad  \epsilon D_R^* = \frac{\lambda_3 - i\lambda_4}{\sqrt{2}} \ ,
\end{equation}
where the $\lambda$s are four independent two component spinors. In this base our condensate becomes simply
\begin{equation}
\langle \lambda_1^2 + \lambda_2^2 +\lambda_3^2 + \lambda_4^2 \rangle \ ,
\end{equation}
which clearly is an $SO(4)$ invariant. Of the original $SU(4)$ global
symmetry we are left with nine broken generators with associated
Goldstone bosons.

In terms of the underlying degrees of freedom, and focusing only on
the techniflavor symmetries, the nine Goldstone bosons transform like
\begin{eqnarray}
 \overline{D}_RU_L \ , \qquad \overline{U}_RD_L \ , \qquad \frac{1}{\sqrt{2}} (\overline{U}_RU_L -\overline{D}_RD_L) \ ,
\end{eqnarray}
for the three which will be eaten by the longitudinal components of
the massive electroweak gauge bosons. The electric charge is
respectively one, minus one and zero. {}For the other six Goldstone
bosons we have
\beq \begin{array}{ccc}
U_L U_L \ , & \quad  D_LD_L  \ , & \quad  U_L D_L \ , \label{othergbs}
\end{array}\eeq
with the following electric charges
\begin{eqnarray}
y +1\ , \qquad y-1 \ ,\qquad y \ ,
\end{eqnarray}
together with the associated anti-particles.
The last six Goldstone bosons (Eq.~(\ref{othergbs})) are
di-technibaryons  with opposite baryonic charge, one and minus one,
respectively. The baryon number is a diagonal generator of
$SU(4)$. As we already mentioned the choice of $y = 1$ makes one
of the Goldstone bosons (namely the $D$) electrically neutral. We
will explore the possibility of a neutral di-technibaryon as a
component of cold dark matter in section \ref{secdarkside}.

\section{Effective Theories}

While the leptonic sector can be described within perturbation
theory since it interacts only via electroweak interactions, the
situation for the techniquarks is more involved since they combine into
composite objects interacting strongly among themselves. It is therefore
useful to construct low energy effective theories encoding the basic
symmetry features of the underlying theory. We construct the linearly
and nonlinearly realized low energy effective theories for our underlying
theory. The theories we will present can be used to investigate
 relevant processes of interest at LHC and LC. It would be interesting to perform the analysis in \cite{Zerwekh:2005wh} with these specific theories.

\subsection{The Linear Realization}
The relevant effective theory for the Higgs sector at the electroweak scale consists, in our model, of a light composite Higgs and nine Goldstone bosons. These
can be assembled in the matrix
\begin{eqnarray}
M = \left(\frac{\sigma}{2} + i\sqrt{2}\Pi^a\,X^a\right)E \ ,
\end{eqnarray}
which transforms under the full $SU(4)$ group according to
\begin{eqnarray}
M\rightarrow uMu^T \ , \qquad {\rm with} \qquad u\in SU(4) \ ,
\end{eqnarray}
and $X^a$ are the generators of the $SU(4)$ group which do not leave invariant the vacuum expectation value of $M$
\begin{eqnarray}
\langle M \rangle = \frac{v}{2}E
 \ .
\end{eqnarray}
It is convenient to separate the fifteen generators of $SU(4)$ into
the six that leave the vacuum invariant ($S^a$) and the other nine
that do not ($X^a$). One can show that the $S^a$ generators of the $SO(4)$
subgroup satisfy the following relation
\begin{eqnarray}
S^a\,E + E\,{S^a}^{T} = 0 \ ,\qquad {\rm with}\qquad  a=1,\ldots  ,  6 \ .
\end{eqnarray}
The explicit realization of the generators is shown in appendix
\ref{appgen}.

The electroweak subgroup can be embedded in $SU(4)$, as explained in
detail in \cite{Appelquist:1999dq}. The main difference here is that
we have a more general definition of the hypercharge. The electroweak
covariant derivative is
\begin{eqnarray}
D_{\mu}M =\partial_{\mu}M - i\,g \left[G_{\mu}M + MG_{\mu}^T\right]  \
, \label{covariantderivative}
\end{eqnarray}
with
\begin{eqnarray}
G_{\mu} = \left(%
\begin{array}{cc}
  W_{\mu} & 0 \\
  0 & -\frac{g^{\prime}}{g}B_{\mu}^T \\
\end{array}%
\right) + \frac{y}{2}\frac{g^{\prime}}{g} B_{\mu}\left(%
\begin{array}{cc}
  1 & 0 \\
  0 & -1 \\
\end{array}%
\right) \ .
\end{eqnarray}
We also have
\begin{eqnarray}
W_{\mu} = W_{\mu}^a\frac{\tau^a}{2}\ , \qquad B_{\mu}^T = B_{\mu}\frac{{\tau^3}^T}{2} =B_{\mu}\frac{{\tau^3}}{2} \ ,
\end{eqnarray}
where $\tau^a$ are the Pauli matrices.
It is convenient to rewrite the gauge bosons in a more compact
form
\begin{eqnarray}
G=W^a\,L^a - \frac{g^{\prime}}{g} B_{\mu}{R^3}^T
+\sqrt{2}{y}\frac{g^{\prime}}{g}B_{\mu} S^4 \ ,
\end{eqnarray}
with
\begin{eqnarray}
L^a = \frac{S^a + X^a}{\sqrt{2}} \ , \qquad {R^a}^T =\frac{X^a -
  S^a}{\sqrt{2}} \ ,\quad {\rm and} \quad  a=1,2,3 \ .
\end{eqnarray}
With this gauging we are ensuring the correct pattern of electroweak
symmetry breaking. In fact we can rewrite
\begin{eqnarray}
G=G_S + G_X \ ,
\end{eqnarray}
with
\begin{eqnarray}
G_S= \frac{1}{\sqrt{2}}\sum_{a=1}^3 S^a \left[W^a +
  \frac{g^{\prime}}{g} B\delta_a^3\right]
+\sqrt{2}{y}\frac{g^{\prime}}{g}B\,S^4 \ ,\qquad G_X=
\frac{1}{\sqrt{2}}\sum_{a=1}^3 X^a \left[W^a - \frac{g^{\prime}}{g}
  B\delta_a^3\right] \ . \quad\, \label{gaugeGSGX}
\end{eqnarray}
The generators satisfy the normalization conditions ${\rm Tr}[X^a\, X^b] = \delta^{ab}/2$, ${\rm Tr}[S^a\, S^b] = \delta^{ab}/2$ and ${\rm Tr}[SX]=0$.
Three of the Goldstone bosons, in the
unitary gauge, are absorbed in the longitudinal degrees of freedom of
the massive weak gauge bosons while the extra six Goldstone bosons will
acquire a mass due to extended technicolor interactions as well as the
electroweak interactions per se. Assuming a bottom up approach we will introduce by hand
a mass term for the Goldstone bosons. The new Higgs
Lagrangian is then
\begin{eqnarray}
L &=& \frac{1}{2}{\rm Tr}\left[D_{\mu}M D^{\mu}M^{\dagger}\right]  +
\frac{m^2}{2}{\rm Tr}[MM^{\dagger}] \nonumber \\&-&\frac{\lambda}{4}
     {\rm Tr}\left[MM^{\dagger} \right]^2 - \widetilde{\lambda} {\rm
       Tr}\left[M M^{\dagger} M M^{\dagger}\right] -\frac{1}{2}\,
     \Pi_a (M^2_{\rm ETC})^{ab} \Pi_b  \ ,
\end{eqnarray}
with $m^2 > 0$ and $a$ and $b$ running over the six uneaten Goldstone
bosons.
The matrix $M^2_{ETC}$ is dynamically generated and
parametrizes our ignorance about the underlying extended technicolor
model yielding the specific mass texture. The pseudo Goldstone bosons are
expected to acquire a mass of the order of a TeV.
Direct and computable contributions from the electroweak corrections
break $SU(4)$ explicitly down to $SU(2)_L\times SU(2)_R$ yielding an
extra contribution to the
uneaten Goldstone bosons. However the main contribution comes from the ETC interactions.

The relation between the vacuum expectation value of the Higgs and the parameters of the present theory is
\begin{eqnarray}
v^2=\langle \sigma \rangle^2 = \frac{m^2}{\lambda + \widetilde{\lambda}} \ .
\end{eqnarray}
{}Since in our theory we expect a light composite Higgs whose mass (in the broken phase) is $2m^2$
\footnote{Note that if one assumes a strongly coupled linear sigma
  model the relation between the physical mass and the mass parameter
  in the theory is no longer linear and important modifications are
  expected \cite{Dietrich:2005jn}.} this corresponds to a small overall self
  coupling. We
have predicted in \cite{Dietrich:2005jn} a Higgs mass in the range
$M_H \simeq 90 -150$ GeV. By choosing the fiducial value $125$~GeV
and
recalling that in our conventions we have $M_{W} = \frac{v\,g}{2}$, we
  then find
\begin{eqnarray}
\lambda + \widetilde{\lambda} \approx \frac{1}{8} \ ,\qquad {\rm with} \qquad v\approx 250~{\rm GeV} \ .
\end{eqnarray}
$\lambda + \widetilde{\lambda}$ corresponds to
the Higgs self coupling in the SM. It turns out that due to the
presence of
a light Higgs the associated sector can be treated
perturbatively. We stress that the expectation of a light composite Higgs relies on
the assumption that the quantum chiral phase transition as function of
number of flavors near the nontrivial infrared fixed point is smooth and
possibly of second order \footnote{We have provided supporting
  arguments for this picture in \cite{Dietrich:2005jn} where the
  reader will find also a more general discussion of this issue and
  possible pitfalls.}. The composite Higgs
Lagrangian is a low energy effective theory and higher dimensional
operators will also be phenomenologically relevant.

\subsection{The Non-Linearly Realized Effective Theory}
One can always organize the
low energy effective theory in a derivative expansion.
The best way is to make use of the exponential map
\beq
U=\exp\left({i\frac{\Pi^a X^a}{F}}\right) E \ ,
\eeq
where $\Pi^a$ represents the 9 Goldstone bosons and $X^a$ are the 9
generators of $SU(4)$ that do not leave the vacuum invariant (see
appendix \ref{appgen} for an explicit realization of the group generators).
To introduce the electroweak interactions one
simply adopts the same covariant derivative used for the linearly realized
effective theory, see
Eqs. (\ref{covariantderivative}-\ref{gaugeGSGX}).

The associated non-linear effective Lagrangian reads
\beq L = \frac{F^2}{2}{\rm Tr}\left[D_\mu UD^\mu U^\dag\right] -
\frac{1}{2}\Pi_a(M^2_{\rm ETC})^{ab}\Pi_b \ . \eeq
Still the mass squared matrix parametrizes our ignorance about the
underlying ETC dynamics.

A common ETC mass for all the pseudo Goldstone bosons carrying baryon
number can be provided by adding the following term to the previous
Lagrangian
\begin{eqnarray}
2C{\rm Tr} \left[UBU^{\dagger}B\right]+{C} =\frac{C}{4F^2} \sum_{i=1}^6 \Pi^i_B \Pi^i_B \ \ ,
\end{eqnarray}
with
\begin{eqnarray}
B= \frac{1}{2\sqrt{2}} \left(
\begin{array}{cc}
\mathbbm{1} & 0 \\
0 & -\mathbbm{1}
\end{array}
\right) \ .
\end{eqnarray}
Dimensional analysis requires $C\propto
\Lambda^6_{TC}/\Lambda^2_{ETC}$.
A similar term can be added to the linearly realized version of our
theory.

It is straightforward to add the vector meson sector to these
theories, which would then allow to repeat the analysis performed in \cite{Zerwekh:2005wh}.

\section{The Dark-Side of the 5$^{\rm th}$ Force\label{secdarkside}}

We now provide a component for cold dark matter within our model. Such a candidate must be electrically
and color neutral and have a mass above the current experimental
exclusion limits.

According to the choice of the hypercharge there are two distinct
possibilities. If we assume the SM-like hypercharge
assignment for the techniquarks and the new lepton family, the
new heavy neutrino can be an interesting dark matter candidate. For
that, it must be made sufficiently stable by requiring no
flavor mixing with the lightest generations and be lighter than the
unstable charged lepton \cite{Dietrich:2005jn}. This possibility is
currently under investigation \cite{Tuominen}. However, we can also
consider another possibility. We can choose the hypercharge
assignment in such a way that one of the pseudo Goldstone bosons
does not carry electric charge. The dynamics providing masses for
the pseudo Goldstone bosons may be arranged in a way that
the neutral pseudo Goldstone boson is the LTB. If
conserved by ETC interactions the technibaryon number protects the
lightest baryon from decaying. Since the mass of the technibaryons
are of the order of the electroweak scale they may constitute
interesting sources of dark matter. Some time ago in a pioneering
work Nussinov \cite{Nussinov:1985xr} suggested that, in analogy
with the ordinary baryon asymmetry in the Universe, a technibaryon
asymmetry is a natural possibility. A new contribution to the
mass of the Universe then emerges due to the presence of the LTB. It
is useful to
compare the fraction of technibaryon
mass $\Omega_{TB}$ to baryon mass $\Omega_{B}$ in the universe
\begin{eqnarray}
\frac{\Omega_{TB}}{\Omega_B} = \frac{TB}{B} \, \frac{m_{TB}}{m_p} \ ,
\label{dmamount}
\end{eqnarray}
where $m_p$ is the proton mass, $m_{TB}$ is the mass of the LTB.
$TB$ and $B$ are the technibaryon and baryon number densities,
respectively.

Knowing the distribution of dark matter in the galaxy earth based experiments can set stringent limits on the physical features of the dominant component of dark matter \cite{Akerib:2005kh}. Such a distribution, however, is not known exactly \cite{Ma:2003cq} and it depends on the number of components and type of dark matter. In order to determine few features of our LTB particle we make the oversimplified approximation in which our LTB constitutes the whole dark matter contribution
to the mass of the Universe. In this limit the previous ratio should be around
$5$ \cite{Lahav:2006qy}. {}By choosing in our model the
hypercharge assignment $y=1$ the lightest neutral Goldstone boson
is the state consisting of the $DD$ techniquarks. The fact that it
is charged under $SU(2)_L$ makes it detectable in Ge detectors
\cite{Bagnasco:1993st}.

It is well known that weak anomalies violate the baryon and the
lepton number. More precisely, weak processes violate $B+L$, while
they preserve $B-L$. Similarly, the weak anomalies violate also
the technibaryon number, since technibaryons couple weakly. The
weak technibaryon-, lepton- and baryon- number violating effects are
highly suppressed at low temperatures while they are enhanced at
temperatures comparable to the critical temperature of the electroweak
phase transition where
sphaleron processes are active (though sphaleron processes only
occur below the scale of the electroweak phase transition)
\cite{Kuzmin:1985mm}.
With $T^{\ast}$ we define the temperature below which the
sphaleron processes cease to be important. This temperature is not
exactly known but it is expected to be in the range between
$150-250$ GeV \cite{Kuzmin:1985mm}.

Following early analysis
\cite{Barr:1990ca,Harvey:1990qw} we have performed a careful computation of
$\Omega_{TB}/\Omega_B$
within our model. There are few differences with respect to the work
in \cite{Barr:1990ca}.
{}For example, our dark matter candidate is not a typical technibaryon
whose mass is uniquely fixed by the underlying technicolor dynamics
but a pseudo Goldstone boson whose mass is set by yet unspecified dynamics.

\begin{figure}[!tbp]
  \begin{center}
    \mbox{
      \subfigure{\resizebox{!}{4.8cm}{\includegraphics{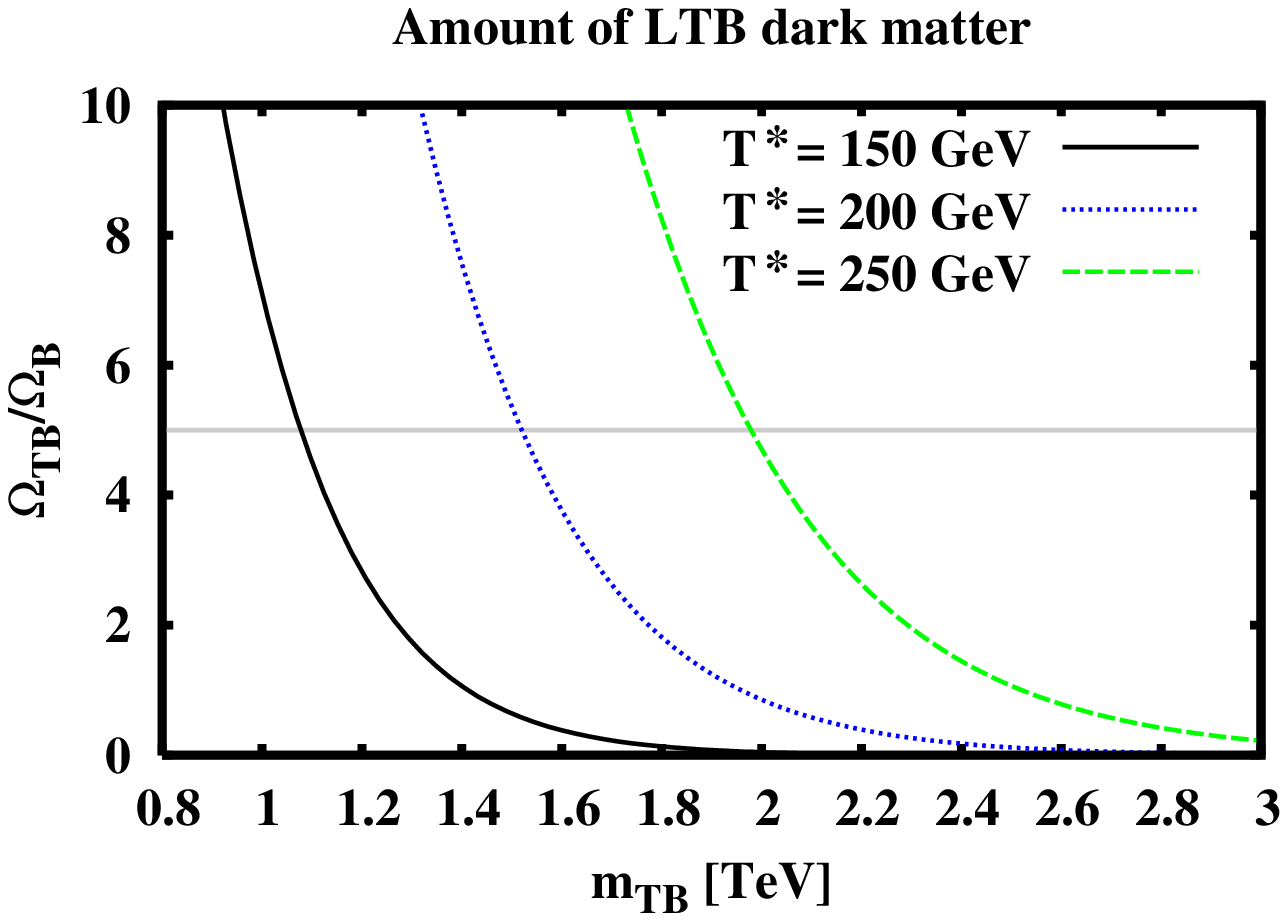}}} \quad
      \subfigure{\resizebox{!}{4.8cm}{\includegraphics{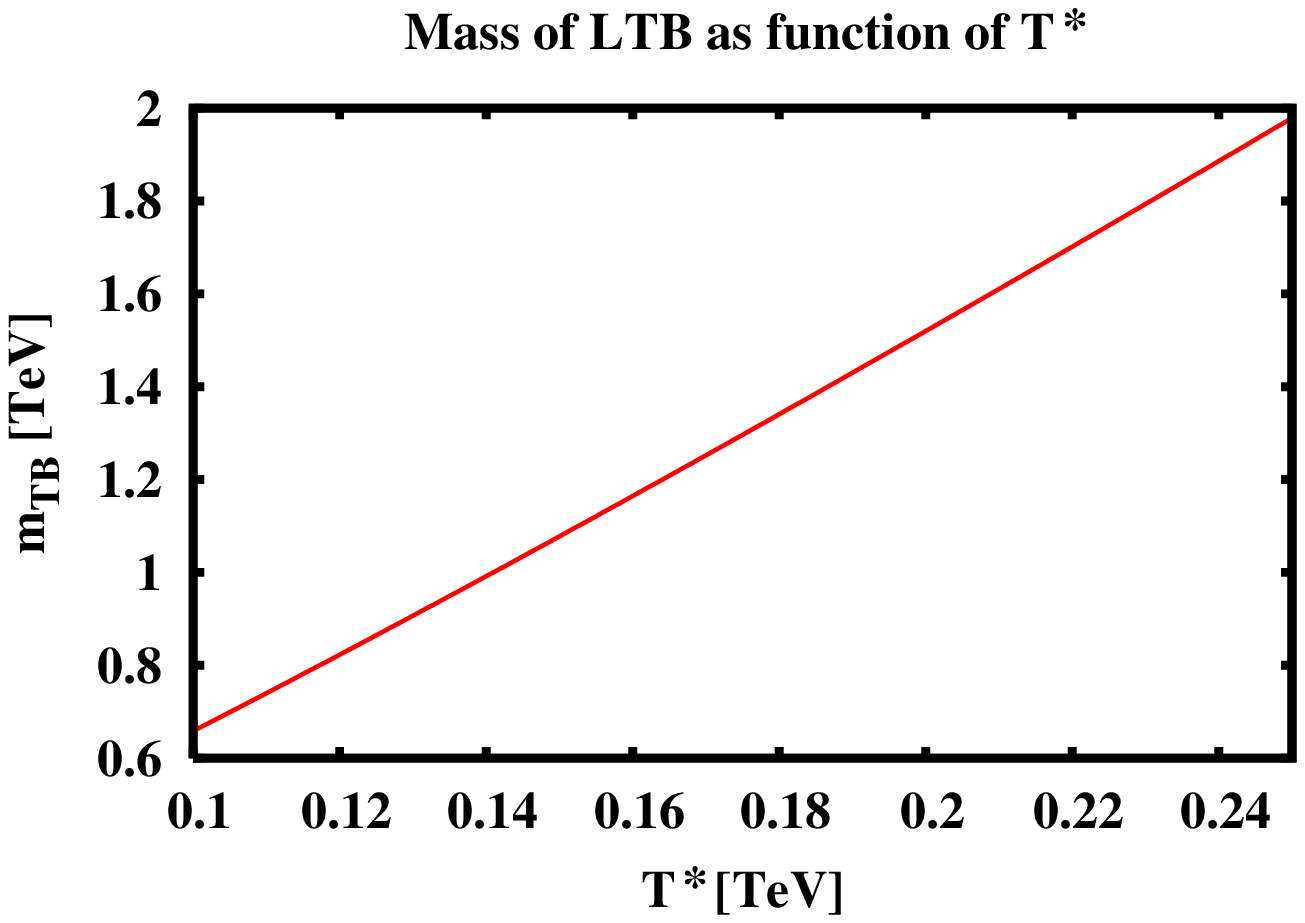}}}
      }
    \caption{Using the values of the parameters indicated in the text:
    \emph{Left Panel}: The fraction of technibaryon matter
    density over the baryonic one as function of the technibaryon
    mass. The desired value of $\Omega_{TB}/\Omega_{B} \sim 5$ depends
    on the lightest technibaryon mass and the value of
    $T^{\ast}$. \emph{Right Panel}: By requiring the correct amount
    ($\Omega_{TB}/\Omega_{B} \sim 5$) of dark
    matter we show the relation between the technibaryon mass and
    $T^{\ast}$.}
    \label{fig:Omega}
  \end{center}
\end{figure}

Imposing thermal equilibrium, electric neutrality condition as well as
the presence of a continuous electroweak phase transition ($T^{\ast}$
now is below the critical temperature) 
we find that the ratio is function of
the technibaryon mass, $T^{*}$, $L/B$ and the new lepton density.
Choosing for simplicity the new lepton density equal to zero and for
display a value of $L/B$  equal to  $-4$, or $-6/7$ if the opposite
sign for TB is chosen,  we get:
\begin{align}
\frac{TB}{B} &=
\frac{11}{36} \, \sigma_{TB}\left(\frac{m_{TB}}{T^\ast}\right) \ ,
\end{align}
with $\sigma_{TB}$ the statistical weight function
\begin{align}
\sigma_{TB}\left(\frac{m_{TB}}{T^\ast}\right) &=
\frac{3}{2\pi^2}\int_0^\infty dx\ x^2
\sinh^{-2}\left(\frac{1}{2}\sqrt{x^2 +
  \left(\frac{m_{TB}}{T^*}\right)^2}\right) \ .
\end{align}
In the previous estimate the LTB is taken to be lighter then the other technibaryons and the new lepton number
is violated. We have, however, considered different scenarios and various limits which will be reported in \cite{Gudnason}.
Our basic results are shown in
Fig.~\ref{fig:Omega}.
The desired value of the dark matter fraction in the Universe can
be obtained for a LTB mass of the order of a TeV for
quite a wide range of values of $T^{\ast}$. The only free parameter in
our analysis is essentially the mass of the LTB which is ultimately
provided by ETC interactions.

\section{Conclusions}

Imminent experiments will shed light on the electroweak breaking
sector of the Standard Model. We have constructed the effective
theories associated to a strong fifth force, of technicolor type,
responsible to the breaking of the electroweak theory. These can be
used for studying signatures and processes which are relevant
for the upcoming experiments such as LHC.
Interestingly
one of the neutral pseudo Goldstone bosons is a natural candidate for
a sizable component of cold dark matter, and we have shown that it is possible to ascribe the whole
dark matter in the Universe to the LTB of our theory, given that it
has a mass of $\sim 1$ TeV.

\acknowledgments
We thank S. Bolognesi, D.D. Dietrich, A. Jokinen, K. Petrov, K. Rajagopal and K. Tuominen  for careful reading of the manuscript, discussions and useful comments.

The work of C.K. and F.S. is supported by the Marie Curie Excellence
Grant under contract MEXT-CT-2004-013510. F.S. is also supported as
Skou fellow of the Danish Research Agency.

\appendix
\section{Generators\label{appgen}}

It is convenient to use the following representation of $SU(4)$
\beq S^a = \begin{pmatrix} \bf A & \bf B \\ {\bf B}^\dag & -{\bf A}^T
\end{pmatrix} \ , \qquad X^i = \begin{pmatrix} \bf C & \bf D \\ {\bf
    D}^\dag & {\bf C}^T \end{pmatrix} \ , \eeq
where $A$ is hermitian, $C$ is hermitian and traceless, $B = -B^T$ and
$D = D^T$. The ${S}$ are also a representation of the $SO(4)$
generators, and thus leave the vacuum invariant $S^aE + ES^T = 0\ $.
Explicitly, the generators read
\beq S^a = \frac{1}{2\sqrt{2}}\begin{pmatrix} \tau^a & \bf 0 \\ \bf 0 &
  -\tau^{aT} \end{pmatrix} \ , \quad a = 1,\ldots,4 \ , \eeq
where $a = 1,2,3$ are the Pauli matrices and $\tau^4 =
\mathbbm{1}$. These are the generators for $SU_V(2)\times U_V(1)$.
\beq S^a = \frac{1}{2\sqrt{2}}\begin{pmatrix} \bf 0 & {\bf B}^a \\
{\bf B}^{a\dag} & \bf 0 \end{pmatrix} \ , \quad a = 5,6 \ , \eeq
with
\beq B^5 = \tau^2 \ , \quad B^6 = i\tau^2 \ . \eeq
The rest of the generators which do not leave the vacuum invariant are
\beq X^i = \frac{1}{2\sqrt{2}}\begin{pmatrix} \tau^i & \bf 0 \\
\bf 0 & \tau^{iT} \end{pmatrix} \ , \quad i = 1,2,3 \ , \eeq
and
\beq X^i = \frac{1}{2\sqrt{2}}\begin{pmatrix} \bf 0 & {\bf D}^i \\
{\bf D}^{i\dag} & \bf 0 \end{pmatrix} \ , \quad i = 4,\ldots,9 \ ,
\eeq
with
\beq\begin{array}{r@{\;}c@{\;}lr@{\;}c@{\;}lr@{\;}c@{\;}l}
D^4 &=& \mathbbm{1} \ , & \quad D^6 &=& \tau^3 \ , & \quad D^8 &=& \tau^1 \ , \\
D^5 &=& i\mathbbm{1} \ , & \quad D^7 &=& i\tau^3 \ , & \quad D^9 &=& i\tau^1
\ .
\end{array}\eeq

The generators are normalized as follows
\beq {\rm Tr}\left[S^aS^b\right] = {\rm Tr}\left[X^aX^b\right] =
\frac{1}{2}\delta^{ab} \ , \qquad {\rm Tr}\left[X^iS^a\right] = 0 \ . \eeq

\end{document}